\documentclass[a4paper,twoside,10pt]{article}

\input{jgrg19.sty}
\def\be{\begin{equation}}
\def\ee{\end{equation}}
\def\bea{\begin{eqnarray}}
\def\eea{\end{eqnarray}}
\newcommand{\eq}[1]{(\ref{#1})}
\newcommand{\half}{\frac12}
\def\({\left (}
\def\){\right )}
\def\[{\left [}
\def\[{\right ]}

\begin{document}
%
\pagestyle{fancy}
\fancyhead{}
  \fancyhead[RO,LE]{\thepage}
  \fancyhead[LO]{B.~Craps}                  
  \fancyhead[RE]{Cosmological singularity problem}    
\rfoot{}
\cfoot{}
\lfoot{}
\label{I3}    
\title{%
  The cosmological singularity problem
}
%
\author{%
  Ben Craps\footnote{Email address: Ben.Craps@vub.ac.be}
}
%
\address{%
Theoretische Natuurkunde, Vrije Universiteit Brussel, and \\
International Solvay Institutes \\
Pleinlaan 2, B-1050 Brussels, Belgium
  }
%
\abstract{
Despite impressive phenomenological successes, cosmological models are
incomplete without an understanding of what happened at the big bang
singularity. Depending on the model, one would like to understand how
appropriate initial conditions were selected at the big bang
singularity, or how a pre-existing contracting universe underwent a
big crunch/big bang transition, if such transitions are possible at
all. In this talk, after an introduction to these questions, an attempt 
is described to study cosmological singularities using the AdS/CFT 
correspondence. A specific model in which asymptotically
AdS initial data evolve into a big crunch singularity is discussed and a dual field
theory description is provided.   
}

\section{Introduction: inflation, open questions, and an alternative}

The last two decades have witnessed enormous progress in our understanding of the composition and evolution of the universe. One of the remaining challenges is to understand how the very early universe reached a nearly homogeneous, nearly flat state with a specific spectrum of density perturbations consistent with present observations.

The most popular explanation is that the very early universe underwent a period of inflation  \cite{I3_inflation}. If one assumes that inflation started and lasted long enough, it is able to explain the flatness and homogeneity of the universe. It also solves the monopole problem. The greatest success of inflation is that ``simple'' (single-field, slow-roll) inflationary models predict nearly scale-invariant, nearly Gaussian adiabatic density perturbations \cite{I3_perturbations}. These are the seeds of large scale structure and are visible as temperature anisotropies in the cosmic microwave background (CMB). 

It is important, though, to ask what predictions inflation makes for upcoming observations. Two of the holy grails of observational cosmology in the coming years are the possible observation of non-gaussianities in the CMB, and of CMB polarization due to primordial gravitational waves (tensor modes). The ``simplest'' inflationary models (single-field, slow-roll, two-derivative) predict that non-gaussianities are too small to be observed. However, other models allow for observable levels of non-gaussianity. Simple field theory models of inflation predict a level of tensor modes that might be observable (though it is not guaranteed). Many other models predict a non-detectable level of tensor modes. So we see that the future predictions of inflation are strongly model-dependent. It would therefore be desirable to understand which specific models (if any) are preferred from a particle physics point of view.

In this context, an important feature of inflationary models is that they need ultraviolet (UV) completions. This can be seen in several ways. First, the slow-roll conditions of simple inflationary models can be destroyed by $(1/M_P^2)$-suppressed quantum gravity corrections to the inflaton potential, where $M_P$ is the Planck mass. Sufficient control over the UV of the theory is needed to compute such terms. Second, in models with observable primordial gravity waves, the inflaton $\phi$ moves over distances in field space much larger than $M_P$ \cite{I3_Lyth}. Terms involving arbitrarily high powers of $\phi/M_P$ are thus important, and their coefficients depend on the UV of the theory. Third, in single field, slow-roll inflation, only higher derivative models can give rise to observable levels of non-gaussianity. But in the presence of some higher derivative terms, one needs to know the UV theory to argue why other higher derivative terms are not present. We conclude that a satisfactory theory of inflation requires a theory beyond general relativity.  
   
In recent years, important progress has been made on constructing inflaton actions in string theory. This is technically not straightforward because it is necessary to stabilize all moduli and to compute all relevant Planckian corrections. See \cite{I3_stringinflation} for a recent review. But even if such effective field theory models with suitable inflaton potentials are found, one may wonder how the universe emerged in a state that allowed inflation to start. In other words, how was a suitably fine-tuned initial state selected? In particular, in general relativity, inflationary solutions are past geodesically incomplete \cite{I3_pastincomplete}. The question should then be asked whether singularity resolution in a more fundamental theory puts constraints on which solutions of the effective theory are allowed.

As we have discussed, even though inflation has had important phenomenological successes, important theoretical questions with phenomenological implications are yet to be answered. This state of affairs has motivated the construction of several alternative models, one of which is the cyclic universe \cite{I3_cyclic}. Inflation (ultra-rapid expansion) is the only known mechanism to dynamically generate the required nearly scale-invariant spectrum of density perturbation in an expanding universe. In a contracting universe, however, the ekpyrotic mechanism (ultra-slow contraction) generates a spectrum of perturbations very similar to that of inflation \cite{I3_ekpyroticperturbations}. In general relativity, the transition from a contracting to an expanding (spatially flat) universe requires going through a singularity \cite{I3_crunchbang}. At present, it is unclear whether such a transition is possible and whether perturbations would go through essentially unchanged. The answer will have to come from a theory beyond general relativity.

We have seen that for the inflationary universe, and even more for alternative models, it is important to try and understand the big bang singularity. The work described in this talk is motivated by several fundamental questions. Can we describe the big bang itself? How do space and time emerge from the big bang? Is it consistent to have a contracting universe before the big bang? Does the universe have a natural initial state, and if so, does it lead to inflation?   

In section~\ref{sec:I3_emergent}, we will first briefly review the concept of emergent space-time as well as the holographic principle. In section~\ref{sec:I3_AdS}, we will discuss the AdS/CFT correspondence and ``AdS cosmologies''. In section~\ref{sec:I3_ABJM}, we will give a slightly more detailed account of specific ($AdS_4$/ABJM) models with cosmological singularities. Section~\ref{sec:I3_conclusions} contains our conclusions and an outlook. 
\section{Emergent space-time and the holographic principle}\label{sec:I3_emergent}

We first briefly review the concept of emergent space-time -- a more elaborate review can be found in \cite{I3_Seiberg_Solvay}. To perform a measurement with desired resolution $\Delta x$, one needs light with wavelength $\lambda<\Delta x$. During the measurement, the energy $E=hc/\lambda$ of a photon gets concentrated in region with dimension $\Delta x$. This energy is within its own Schwarzschild radius if $\Delta x$ is smaller than the Planck length $\ell_P=(hG/c^3)^{1/2}\approx 10^{-35}\,m$. If one attempts to probe space with a resolution smaller than $\ell_P$, one creates a black hole with size larger than $\ell_P$. As a consequence, the concept ``space'' loses its meaning below the Planck length.

An emergent concept in physics is absent in the fundamental description of a theory, but appears in a macroscopic description. Examples include the classical path of a particle, which has no fundamental meaning in quantum mechanics, but emerges as a good approximation for macroscopic objects, and the description of water as a continuous fluid, which is only valid on scales greater than the mean free path of the water molecules.

Since space loses its meaning near the Planck scale, a natural idea is that space may be an emergent concept. This idea raises many questions, though. If space only plays a role in a ``macroscopic'' description of reality, then what is the fundamental, underlying theory? In other words, what are the ``molecules'' in the analogy with water? How exactly does space emerge on ``macroscopic'' scales? In particular, how does space originate in cosmology? And if space is emergent, then how about time?

A framework in which the emergence of space has been realized is the holographic principle. In optics, holography means that 3d information is completely encoded on a 2d photographic plate and can be reconstructed starting from the 2d image. In quantum gravity, the holographic principle \cite{I3_holographic} says, roughly speaking, that gravitational physics in a space-time is exactly equivalent to a non-gravitational theory on the boundary of this space-time. 

The motivation for the holographic principle goes back to studies of black hole entropy. Since the number of bits of information in a black hole is proportional to the area of the horizon rather than the volume inside the horizon \cite{I3_BH}, maybe the carriers of the information live on the horizon rather than in the interior.  

\section{The AdS/CFT correspondence and AdS cosmologies}\label{sec:I3_AdS}

In the second half of the 90s, the holographic principle has been realized rather precisely in the context of string theory. String theory is a proposed framework for force unification, in which elementary particles are described as small oscillating strings. The theory automatically contains gravity and it is consistent with the rules of quantum mechanics, making it a model of quantum gravity. Important features of the theory are that, in the simplest models, space-time has more than four dimensions, and that, in addition to strings, the theory contains extended objects known as branes.

We will mention two string theory realizations of the holographic principle and then focus on one of them. One realization is matrix theory \cite{I3_BFSS}, a quantum mechanical theory of large matrices. At large distances, the model is well-approximated by gravity in a space-time (the analogue of water as a continuous fluid). On the other hand, at small distances, the description in terms of space-time no longer applies -- the notion of space is replaced by non-commuting matrices (the analogue of the molecular structure). The second realization, which will be the focus of the remainder of this talk, is the AdS/CFT correspondence, which relates string theory in asymptotically anti-de Sitter (AdS) space-times to conformal field theories on the conformal boundary of AdS \cite{I3_AdS_CFT}.

The four-dimensional $AdS_4$ space (in global coordinates) has a time coordinate $t$, a radial coordinate $r$ and angular coordinates $\theta$ and $\phi$. The conformal boundary consists of the time direction (labeled by $t$) multiplied by the two-sphere (labeled by $\theta, \phi$). One can therefore wonder how the radial position $r$ in AdS is encoded in the dual field theory. It turns out that the radial position $r$ of an object in AdS corresponds to the energy of the corresponding object in the field theory. The radial direction is emergent.

The hope is now that a dual field theory description will enable us to describe gravitational physics in regimes where gravity breaks down, such as near black hole or cosmological singularities -- the latter will be the focus of the remainder of this talk. Our starting point is the existence of supergravity solutions in which smooth, asymptotically AdS initial data evolve to a big crunch singularity, by which we mean a spacelike singularity that reaches the boundary in finite global time \cite{I3_HH}. Such solutions are called ``AdS cosmologies''. The question is whether a dual gauge theory can be used to study the singularity in quantum gravity. 

To be a little bit more specific, to define (super-)gravity in AdS, one has to specify boundary conditions at spatial infinity (since light signals can reach the boundary in finite time). With the usual, supersymmetric boundary conditions, AdS is perfectly stable and a big crunch is not generated from smooth, asymptotically AdS initial data. This changes, however, when certain modified, non-supersymmetric boundary conditions are imposed \cite{I3_HH}. The modified boundary conditions in the bulk are reflected in the dual field theory by the presence of a potential unbounded from below. This potential is so steep that it causes an operator to reach infinity in finite time. Therefore, the goal becomes to learn something about cosmological singularities (in the bulk) by studying unbounded potentials (in the boundary theory).

In quantum mechanics (as opposed to quantum field theory), steep potentials unbounded from below are well-understood. A priori, the danger is that wavepackets disappear at infinity in finite time, causing a violation of unitarity. This problem can be cured by considering a self-adjoint extension of the Hamiltonian \cite{I3_unbounded}. This amounts to restricting the space of allowed wavefunctions so that the Hamiltonian becomes self-adjoint and time evolution therefore unitary. The simplest choices of self-adjoint extensions can be visualized as putting a ``brick wall'' boundary condition at infinity, such that a ``mirror'' wavepacket appears when the original wavepacket disappears, thereby conserving probability. There also exist consistent self-adjoint extensions, though, where a wavepacket disappears at one boundary at infinity and reappears at another boundary, again preserving total probability. While well-understood in quantum mechanics, the extension of the method of self-adjoint extensions to quantum field theory is highly non-trivial -- a first attempt appeared in \cite{I3_CHT1} and an update will appear elsewhere \cite{I3_CHT3}. 
   
\section{Cosmological singularities in $AdS_4$/ABJM}\label{sec:I3_ABJM}

Eleven-dimensional supergravity on $S^7/\mathbb{Z}_k$ allows a consistent truncation to four-dimensional gravity coupled to a single scalar field $\varphi$. The potential has a maximum for vanishing scalar field that corresponds to the $AdS_4$ vacuum solution. Small fluctuations around the the AdS solution have a mass in the range $m_{BF}^2<m^2<m_{BF}^2+1$. Since the Breitenlohner-Freedman bound \cite{I3_Breitenlohner82} is satisfied, the maximally supersymmetric solution, with the standard boundary conditions, is both perturbatively and non-perturbatively stable. 
In global coordinates, asymptotically $AdS_4$ metrics satisfy
\be
ds^2 \sim -(1+ r^2) dt^2 + \frac{dr^2}{1+r^2} + r^2 d\Omega_2^2\ \ \ \ (r\to\infty)\,.\label{eqn:I3_globalAdS4}
\ee
In any asymptotically AdS solution, the scalar field behavior at large radial coordinate is
\be\label{eqn:I3_varphi4}
\varphi(t,r,\Omega) \sim \frac{\alpha(t,\Omega)}{r} +\frac{\beta(t,\Omega)}{r^2}\ \ \ \ (r\to\infty)\,.
\ee 
The standard, supersymmetric boundary conditions correspond to $\beta=0$. There exists however a whole one-parameter family of AdS invariant boundary conditions,
\be\label{eqn:I3_BCs4}
\beta= - h \alpha^2\,,
\ee
where $h$ is an arbitrary constant \cite{I3_Hertog:2004dr}. For $h \neq 0 $, smooth asymptotically AdS initial data can evolve into a big crunch singularity \cite{I3_HH}.

M-theory in asymptotically $AdS_4\times S^7/\mathbb{Z}_k$ space-times with $\beta =0$ boundary conditions is dual to the three-dimensional superconformal field theory that describes the low energy dynamics of coincident M2-branes. In \cite{I3_ABJM}, Aharony, Bergman, Jafferis and Maldacena (ABJM) proposed a specific three-dimensional $\mathcal N=6$ superconformal $U(N) \times U(N)$ Chern-Simons-matter theory with levels $k$ and $-k$ as the world-volume theory of $N$ coincident M2-branes on a $\mathbb{C}^4 / \mathbb{Z}_k$ singularity. Here, $\mathbb{Z}_k$ is generated by $y^A\to\exp(2\pi i/k)\,y^A$. Besides the two $U(N)$ gauge fields $A$ and $\hat A$, the theory contains scalar fields $Y^A$, $A=1,\dots,4$, transforming in the fundamental representation of the $SU(4)_R$ R-symmetry group and in the bifundamental $(N, \bar N)$ of the gauge group, as well as fermions. The scalars come with a sextic single trace potential (which can be thought of as the analogue of the commutator squared potential of ${\cal N}=4$ super-Yang-Mills theory). The ABJM theory has a 't~Hooft limit $N\to\infty$ with $N/k$ fixed; in this limit, one actually finds a weakly coupled type IIA string theory. The operator dual to the bulk scalar $\varphi$ is the dimension one chiral primary operator \cite{I3_CHT2}
\be
\mathcal O \sim Tr \(Y^1 Y_1^{\dagger} - Y^2Y_2^{\dagger} \). \label{eqn:I3_O4}
\ee 

The boundary condition \eq{eqn:I3_BCs4} corresponds to adding a classically marginal triple trace deformation to the boundary action 
\be\label{eqn:I3_hdef}
S = S_{ABJM} + {\rm conf.\ coupl.}\ + \frac{h}{N^4} \int d^3 x \left[ Tr \(Y^1 Y_1^{\dagger} - Y^2Y_2^{\dagger} \) \right]^3 \,.
\ee
The second term is the conformal coupling of the scalars due to the curvature of $S^2$. Two things to note about this deformation of the ABJM theory are that it is conformal in the planar limit, and that it reduces to the $O(2N^2)\times O(2N^2)$ vector model in the weak coupling limit $N\to\infty,\ N/k\to 0$ \cite{I3_CHT2}. 

Recently, a brane interpretation of the instability of \eq{eqn:I3_hdef} has been provided in \cite{I3_BC}. Consider spherical M2-branes in $AdS_4\times S^7$, extended along the time direction and a two-sphere inside $AdS_4$ and localized at a point in $S^7$. Such branes could either nucleate out of empty $AdS_4\times S^7$ or be present in an initial state. Now compute the effective potential for the radius $R$ of such branes, or rather, for the canonically normalized field $\phi\equiv\sqrt R$. With the standard supersymmetric boundary conditions, cancellations related to supersymmetry leave behind a quadratic potential for $\phi$, which precisely corresponds to the conformal coupling terms in \eq{eqn:I3_hdef}. If a spherical brane nucleates, this potential causes it to shrink again and disappear. With the modified boundary conditions \eq{eqn:I3_BCs4}, however, an unbounded term sextic in $\phi\equiv\sqrt R$ is added to the potential, in agreement with the triple trace term in \eq{eqn:I3_hdef}. The presence of this term implies that a sufficiently large spherical M2-brane will be pulled to infinite radius in finite time. M2-branes are domain walls in $AdS_4$: the 4-form flux inside a spherical brane is one unit smaller than outside. In a process in which spherical branes nucleate and grow, the 4-form flux $N_{eff}$ at a fixed radial position $r$ decreases with time. As a consequence, the effective 't~Hooft coupling $N_{eff}/k$ decreases as a function of time. Since small 't~Hooft coupling corresponds to large curvature in the bulk, the result in a gravity approximation will be a curvature singularity, namely the big crunch singularity visible in the supergravity solutions describing AdS cosmologies.\footnote{I would like to thank Y.~Nakayama for a related discussion.}

To understand the dual description of our AdS cosmology, we need to study the dynamics of the field theory defined by \eq{eqn:I3_hdef}. For now, we will do so in the weakly coupled 't~Hooft limit $N\to\infty,\ N/k\to 0$, in which the model reduces to the $O(2N^2)\times O(2N^2)$ vector model. Let us therefore first review the well-studied $O(N)$ vector model, defined by the action
\be
S=\int d^3x\left(
-\half\partial_\mu\vec\phi\cdot\partial^\mu\vec\phi-\frac{\lambda}{6N^2}\left(\vec\phi\cdot\vec\phi\right)^3 \right).
\ee
Its perturbative beta function is given by \cite{I3_Pisarski}
\be
\beta_{\rm pert}(\lambda)=\frac{3}{2\pi^2N}\left(\lambda^2-\frac{\lambda^3}{192}\right) + \ {\rm higher\ order\ in}\ \frac{1}{N}.
\ee
For negative values of the coupling ($\lambda<0$), there is a UV fixed point at $\lambda=0$, so that the model is asymptotically free (though of course it has a potential unbounded from below) \cite{I3_CG}. For positive coupling ($\lambda>0$), the model exhibits a perturbative fixed point at $\lambda^*=192$ \cite{I3_Pisarski}. There is, however, a non-perturbative instability at leading order in the $1/N$ expansion whenever $\lambda>\lambda_c\equiv 16\pi^2$ \cite{I3_BMB}; the perturbative fixed point lies in the unstable regime. Working with Poincar\'e invariant states, one can show that the effective potential approaches minus infinity as the renormalized value of $\langle\phi^2\rangle$ goes to minus infinity. (If one introduces a cutoff, the effective potential is bounded from below, but all masses are of the order of the cutoff, so that there is no interesting continuum limit.) Recently, time-dependent states have been considered in this model. It has been found that states with 
\be
\langle \phi^2 \rangle_{\rm ren}=-\frac{CN}{t}
\ee  
exist at least up to some time $t=0$ \cite{I3_ARS}.

We can summarize the analysis of the $O(N)$ vector model as follows. There is a classical instability for $\lambda<0$, in which case $\phi\sim 1/\sqrt{|t|}$. There is a quantum instability for $\lambda>\lambda_c$, in which case $\langle\phi^2 \rangle_{\rm ren}\sim 1/|t|$. The perturbative fixed point lies in the quantum unstable regime.

Now let us consider the $O(N)\times O(N)$ vector model \cite{I3_RSB} (where in the application we have in mind, $N$ will be replaced by $2N^2$):
\bea
S&=&\int d^3x\left[
-\half\partial_\mu\vec\phi_1\cdot\partial^\mu\vec\phi_1-\half\partial_\mu\vec\phi_2\cdot\partial^\mu\vec\phi_2-\frac{\lambda_{111}}{6N^2}\left(\vec\phi_1\cdot\vec\phi_1\right)^3 - \frac{\lambda_{222}}{6N^2}\left(\vec\phi_2\cdot\vec\phi_2\right)^3\right.\cr
&&\left.\ \ \ \ \ \ \ \ \ \  - \frac{\lambda_{112}}{6N^2}\left(\vec\phi_1\cdot\vec\phi_1\right)^2\left(\vec\phi_2\cdot\vec\phi_2\right)- \frac{\lambda_{122}}{6N^2}\left(\vec\phi_1\cdot\vec\phi_1\right)\left(\vec\phi_2\cdot\vec\phi_2\right)^2
\right].
\eea
As a special case, we recover the potential corresponding to the last term in \eq{eqn:I3_hdef},
\be\label{eqn:I3_V}
V=\frac{\lambda}{6N^2}\left(\vec\phi_1\cdot\vec\phi_1-\vec\phi_2\cdot\vec\phi_2\right)^3,
\ee
where by convention we choose $\lambda<0$ (the other choice would correspond to interchanging the roles of $\vec\phi_1$ and $\vec\phi_2$). The results of the analysis in \cite{I3_CHT2} are as follows. If one starts with the potential \eq{eqn:I3_V} and follows the four independent sextic couplings under perturbative renormalization group flow towards the UV, one approaches a UV fixed point with $\lambda_{222}=\lambda^*,\ \lambda_{112}=\lambda_{122}=\lambda_{111}=0$ -- the approach of the fixed point has also been computed. However, as in the $O(N)$ vector model, a quantum instability will kick in before $\lambda_{222}$ reaches $\lambda^*$.

Taking into account this quantum instability, the dynamics is as follows. Start with a small negative value of $\lambda$, so that $\vec\phi_1$ is classically unstable. As $\vec\phi_1$ rolls to large values, a lot of energy becomes available, so one needs to run the couplings towards the UV. At some time, the coupling $\lambda_{222}$ becomes larger than $\lambda_c$, so that $\vec\phi_2$ becomes quantum unstable. The coupled system of $\vec\phi_1$ and $\vec\phi_2$ needs to be studied -- this is work in progress \cite{I3_CHT3}. Once the approach of the singularity ($t=0$) is understood, it will be interesting to study if and how time evolution can be extended beyond $t=0$.

\section{Conclusions and outlook}\label{sec:I3_conclusions}
In this talk, we have argued that cosmological models are incomplete without an understanding of the cosmological singularity. Then we have seen that the holographic principle maps difficult questions in gravity to (hopefully simpler) questions in quantum field theory. In particular, the AdS/CFT correspondence relates gravitational theories allowing big crunch singularities to field theories with potentials unbounded from below. Therefore, it is interesting to study the dynamics of field theories with unbounded potentials. We have seen preliminary results in a concrete model, namely a deformation of ABJM theory.

What can we hope these models will lead to? In principle, one would like to carry out the following program. Start with a state in the bulk theory (with modified boundary conditions) corresponding to a large, asymptotically AdS space-time with some profile for the scalar field. Translate this state, using the AdS/CFT correspondence, to a state in the dual field theory on the boundary (with a steep unbounded potential). In the dual field theory, evolve the state through the singularity using a self-adjoint extension (if a consistent and natural self-adjoint extension exists). Finally, translate the evolved state back to a state in the bulk theory, and ask whether it has a geometric interpretation. If the boundary theory described only homogeneous modes, experience with self-adjoint extensions in quantum mechanics would suggest that the final state would roughly resemble the initial state, which would suggest a cosmological bounce. Inhomogeneous modes can drastically change this picture, though: particle creation can be potentially attractive for cosmology, but one needs to make sure that backreaction is sufficiently small for the computations to be reliable. In fact, self-adjoint extensions of ``brick wall'' type tend to lead to too much particle creation. Preliminary results suggest that other boundary conditions may be better behaved \cite{I3_CHT3}.

\section*{Acknowledgements}
I am grateful to the organizers of JGRG19, in particular Kei-ichi Maeda, for their kind invitation to this stimulating workshop. I would also like to thank Alice Bernamonti, Thomas Hertog and Neil Turok for collaboration on the results presented in this talk. This research is supported in part by the Belgian Federal Science Policy Office through the Interuniversity Attraction Pole IAP VI/11 and by FWO-Vlaanderen through projects G.0428.06 and G011410N. 


\begin{thebibliography}{99}
%
\bibitem{I3_inflation}
A.~H.~Guth, Phys.\ Rev.\  D {\bf 23} (1981) 347;
A.~D.~Linde, Phys.\ Lett.\  B {\bf 108} (1982) 389;
A.~J.~Albrecht and P.~J.~Steinhardt, Phys.\ Rev.\ Lett.\  {\bf 48} (1982) 1220.
%
\bibitem{I3_perturbations}
A.~H.~Guth and S.~Y.~Pi, Phys.\ Rev.\ Lett.\  {\bf 49} (1982) 1110;
J.~M.~Bardeen, P.~J.~Steinhardt and M.~S.~Turner, Phys.\ Rev.\  D {\bf 28} (1983) 679;
S.~W.~Hawking, Phys.\ Lett.\  B {\bf 115} (1982) 295;
A.~A.~Starobinsky, Phys.\ Lett.\  B {\bf 117} (1982) 175.
%
\bibitem{I3_Lyth}
D.~H.~Lyth, Phys.\ Rev.\ Lett.\  {\bf 78} (1997) 1861 [arXiv:hep-ph/9606387].
%
\bibitem{I3_stringinflation}
D.~Baumann and L.~McAllister, Ann.\ Rev.\ Nucl.\ Part.\ Sci.\  {\bf 59} (2009) 67
[arXiv:0901.0265 [hep-th]].
%
\bibitem{I3_pastincomplete}
A.~Borde, A.~H.~Guth and A.~Vilenkin, Phys.\ Rev.\ Lett.\  {\bf 90}, 151301 (2003)
[arXiv:gr-qc/0110012].
%
\bibitem{I3_cyclic}
P.~J.~Steinhardt and N.~Turok, arXiv:hep-th/0111030.
%
\bibitem{I3_ekpyroticperturbations}
J.~Khoury, B.~A.~Ovrut, P.~J.~Steinhardt and N.~Turok, Phys.\ Rev.\  D {\bf 66} (2002) 046005
[arXiv:hep-th/0109050].
%
\bibitem{I3_crunchbang}
J.~Khoury, B.~A.~Ovrut, N.~Seiberg, P.~J.~Steinhardt and N.~Turok, Phys.\ Rev.\  D {\bf 65} (2002) 086007
[arXiv:hep-th/0108187].
%
\bibitem{I3_Seiberg_Solvay}
N.~Seiberg, arXiv:hep-th/0601234.
%
\bibitem{I3_holographic}
G.~'t Hooft, arXiv:gr-qc/9310026;
L.~Susskind,  J.\ Math.\ Phys.\  {\bf 36}, 6377 (1995)
[arXiv:hep-th/9409089].
%
\bibitem{I3_BH}
J.~D.~Bekenstein, Phys.\ Rev.\  D {\bf 7} (1973) 2333;
S.~W.~Hawking, Commun.\ Math.\ Phys.\  {\bf 43} (1975) 199 [Erratum-ibid.\  {\bf 46} (1976) 206].
%
\bibitem{I3_BFSS}
T.~Banks, W.~Fischler, S.~H.~Shenker and L.~Susskind, Phys.\ Rev.\  D {\bf 55}, 5112 (1997)
[arXiv:hep-th/9610043].
%
\bibitem{I3_AdS_CFT}
J.~M.~Maldacena, Adv.\ Theor.\ Math.\ Phys.\  {\bf 2}, 231 (1998) [Int.\ J.\ Theor.\ Phys.\  {\bf 38}, 1113 (1999)]
[arXiv:hep-th/9711200];
 S.~S.~Gubser, I.~R.~Klebanov and A.~M.~Polyakov, Phys.\ Lett.\  B {\bf 428}, 105 (1998)
[arXiv:hep-th/9802109].
 E.~Witten, Adv.\ Theor.\ Math.\ Phys.\  {\bf 2}, 253 (1998)
 arXiv:hep-th/9802150].
%
\bibitem{I3_HH}
T.~Hertog and G.~T.~Horowitz, JHEP {\bf 0407}, 073 (2004)
[arXiv:hep-th/0406134].
%
\bibitem{I3_unbounded}
M.~Reed and B.~Simon, ``Methods Of Modern Mathematical Physics. 2. Fourier Analysis, Selfadjointness,'' {\it  New York 1975, 361p};
M.~Carreau, E.~Fahri, S.~Gutmann, P.~F.~Mende, Ann. Phys. {\bf 204} (1990) 186.
%
\bibitem{I3_CHT1}
B.~Craps, T.~Hertog and N.~Turok, arXiv:0712.4180 [hep-th].
%
\bibitem{I3_CHT3}
B.~Craps, T.~Hertog, N.~Turok, in progress. 
%
\bibitem{I3_Breitenlohner82}
P.~Breitenlohner and D.~Z.~Freedman, 
Annals Phys.\  {\bf 144} (1982) 249.
%
\bibitem{I3_Hertog:2004dr}
 T.~Hertog and K.~Maeda,
 JHEP {\bf 0407} (2004) 051
 [arXiv:hep-th/0404261].
%
\bibitem{I3_ABJM}
O.~Aharony, O.~Bergman, D.~L.~Jafferis and J.~Maldacena, JHEP {\bf 0810}, 091 (2008)
[arXiv:0806.1218 [hep-th]].
%
\bibitem{I3_CHT2}
 B.~Craps, T.~Hertog and N.~Turok, Phys.\ Rev.\  D {\bf 80}, 086007 (2009)
[arXiv:0905.0709 [hep-th]].
%
\bibitem{I3_BC}
A.~Bernamonti and B.~Craps, JHEP {\bf 0908}, 112 (2009)
[arXiv:0907.0889 [hep-th]].
%
\bibitem{I3_Pisarski}
R.~D.~Pisarski, Phys.\ Rev.\ Lett.\  {\bf 48} (1982) 574.
%
\bibitem{I3_CG}
S.~R.~Coleman and D.~J.~Gross, Phys.\ Rev.\ Lett.\  {\bf 31}, 851 (1973).
%
\bibitem{I3_BMB}
W.~A.~Bardeen, M.~Moshe and M.~Bander, Phys.\ Rev.\ Lett.\  {\bf 52}, 1188 (1984).
%
\bibitem{I3_ARS}
V.~Asnin, E.~Rabinovici and M.~Smolkin, JHEP {\bf 0908}, 001 (2009)
[arXiv:0905.3526 [hep-th]]; B.~Craps, T.~Hertog, N.~Turok, in progress.
%
\bibitem{I3_RSB}
E.~Rabinovici, B.~Saering and W.~A.~Bardeen, Phys.\ Rev.\  D {\bf 36} (1987) 562.

\end{thebibliography}
\end{document}